# Unusual normal and superconducting state properties observed in hydrothermal $Fe_{1-\delta}Se$ flakes

Shaobo Liu[1,2], Sheng Ma[1,2], Zhaosheng Wang[3], Wei Hu[1,2], Zian Li[1], Qimei Liang[3], Hong Wang[1,2], Yuhang Zhang[1,2], Zouyouwei Lu[1,2], Jie Yuan[1,4,5], Kui Jin[1,2,4,5], Jian-Qi Li[1,2], Li Pi[3], Li Yu[1,2,5], Fang Zhou[1,2,5*], Xiaoli Dong[1,2,4,5], and Zhongxian Zhao[1,2,4,5]

[1] *Beijing National Laboratory for Condensed Matter Physics and Institute of Physics, Chinese Academy of Sciences, Beijing 100190, China*
[2] *School of Physical Sciences, University of Chinese Academy of Sciences, Beijing 100049, China*
[3] *Anhui Province Key Laboratory of Condensed Matter Physics at Extreme Conditions, High Magnetic Field Laboratory of the Chinese Academy of Sciences, Hefei 230031, Anhui, China*
[4] *Key Laboratory for Vacuum Physics, University of Chinese Academy of Sciences, Beijing 100049, China*
[5] *Songshan Lake Materials Laboratory, Dongguan, Guangdong 523808, China*

\* Correspondence to: fzhou@iphy.ac.cn (F.Z.)

## Abstract

The electronic and superconducting properties of $Fe_{1-\delta}Se$ single-crystal flakes grown hydrothermally are studied by the transport measurements under zero and high magnetic fields up to 38.5 T. The results contrast sharply with those previously reported for nematically ordered FeSe by chemical-vapor-transport (CVT) growth. No signature of the electronic nematicity, but an evident metal-to-nonmetal crossover with increasing temperature, is detected in the normal state of the present hydrothermal samples. Interestingly, a higher superconducting critical temperature $T_c$ of 13.2 K is observed compared to a suppressed $T_c$ of 9 K in the presence of the nematicity in the CVT FeSe. Moreover, the upper critical field in the zero-temperature limit is found to be isotropic with respect to the field direction and to reach a higher value of ~42 T, which breaks the Pauli limit by a factor of 1.8.

How the normal-state electronic property influences the superconductivity is an important issue regarding the underlying physics of unconventional iron-based superconductors. Compared to the antiferromagnetic iron arsenides,[1,2] the simplest iron selenide superconductor FeSe is unique because of the presence of a nematic electronic order but no magnetic order.[3,4] The electronic nematicity is associated with a tetragonal-to-orthorhombic structural transition upon cooling to $T_s$ ~ 90 K.[3,4] The superconductivity coexists with the electronic nematicity and shows a relatively low critical temperature $T_c$ ~ 9 K [3-5] in the bulk samples. Most recently, it has been suggested that the physics in the tetragonal background is fundamental for the origin of the unconventional superconductivity.[6] In the FeSe crystals grown by chemical-vapor-transport (CVT) method, the normal-state resistivity $\rho(T)$ displays a metallic behavior, and the structural transition manifests itself in $\rho(T)$ as a clear kink at $T_s$. At $T > T_s$, the Hall resistivity $\rho_{xy}(B)$ has been shown to be linear in field $B$ in accordance with a compensated metal character.[7-9] Below $T_s$, a significant nonlinearity in $\rho_{xy}(B)$ develops and the Hall coefficient $R_H(T)$ drops substantially, followed by its sign change from positive to negative. This has been attributed to the emergence of a high-mobility small electron pocket in the nematic regime.[7,8]

In this work, however, distinctly different electronic and superconducting properties are observed in our single-crystal flakes of Fe$_{1-\delta}$Se grown hydrothermally. First, a metal-nonmetal crossover in the normal state is evident from the resistance measurements. Correspondingly, the resistive kink related to the structural/nematic transition as seen in CVT FeSe is absent here. Moreover, the Hall measurements also imply a disappearance of the additional high-mobility electron band. Interestingly, furthermore, the superconducting $T_c$ (13.2 K) of the hydrothermal Fe$_{1-\delta}$Se is observed to be higher than that (9 K) of the nematic CVT FeSe. Therefore, our results suggest that the emergent high-mobility electron pocket in the nematic FeSe does not favor the superconductivity. This is in agreement with our recent results obtained in isovalently substituted[6] and iron-deficient[10] FeSe single crystals. In addition, compared to the anisotropic upper critical field $H_{c2}(0)$ of CVT FeSe,[11-15] $H_{c2}(0)$ of the present hydrothermal Fe$_{1-\delta}$Se is found not only to be isotropic but also reach a higher value of ~ 42 T, which breaks the BCS Pauli limit [$H_P^{BCS}(0)$ =1.84$T_c$] by a factor of 1.8. An isotropic, high upper critical field is desirable for superconductivity application.[16]

The Fe$_{1-\delta}$Se flakes studied here were obtained by cleaving the samples synthesized using a hydrothermal method similar to our previous work.[17,18] The x-ray diffraction (XRD) experiments were carried out at room temperature on a diffractometer (Rigaku SmartLab, 9 kW) equipped with two Ge (220) monochromators. The scanning electron microscope (SEM) and energy dispersive x-ray (EDX) measurements were carried out on a Hitachi S-4800. The experiments of transmission electron microscopy (TEM) and selected-area electron diffraction (SAED) were performed at room temperature on an electron microscope (TEM, ARM200F, JEOL Ltd) equipped with a spherical aberration corrector (CEOS GmbH). The electrical transport properties under magnetic fields up to 9 T were measured with a six-probe configuration on a Quantum Design PPMS-9 system. The magnetotransport data under high fields up to 38.5 T were measured with a current $I = 1$ mA at temperatures down to 1 K in the High Magnetic Field Laboratory of CAS. The sample dimensions for the transport measurements were 1.5 mm in length, 0.6 mm in width and 400–800 nm in thickness. The superconducting diamagnetic signals were measured on a Quantum Design MPMS-XL1 system with a tiny remnant field < 4 mOe.

The Fe$_{1-\delta}$Se flake samples exhibit a single preferred (001) orientation, as demonstrated by the representative XRD pattern shown in Fig. 1a. The SAED pattern taken along the [00l] zone axis is presented in Fig. 1b. Note that no extra spots, but only the tetragonal main spots, are detected by SAED. Both the patterns agree with the space group of *P*4/*nmm*. The lattice constants are obtained as $a \sim 3.75$ Å and $c \sim 5.528$ Å from the SAED and XRD patterns, respectively. The SEM image in the inset of Fig. 1c reveals a smooth surface morphology of the Fe$_{1-\delta}$Se flake. As shown in Fig. 1d, the double-crystal x-ray rocking curve gives a full width at half maximum (FWHM) of 0.45° for the (004) Bragg reflection, indicating a high sample quality. The experiment of EDX spectroscopy (Fig. 1c) detects no impurities in the hydrothermal Fe$_{1-\delta}$Se. The superconducting transition temperature is characterized as $T_c = 13.2$ K by the dc diamagnetic measurement, as shown in Fig. 2a. The onset zero-resistance temperature is ~12 K, as seen from Figs. 2b, 3a and b. The cell parameter $c$ (with some uncertainty only in the third decimal), the pristine SAED pattern, and the $T_c$ value all point to a small non-stoichiometry (with $\delta$ closer to 0) for the hydrothermal Fe$_{1-\delta}$Se, when comparing with previous results.[10,19]

Now we discuss the unusual normal-state transport properties observed in our hydrothermal $Fe_{1-\delta}Se$, with emphasis on their significance in comparison with those seen in nematically ordered CVT FeSe. It is well known that an obvious kink at the structural $T_s \sim 90$ K, as well as a metallic behavior in a wide temperature range, is a characteristic of the temperature-dependent resistivity of the nematic FeSe. However, as can be seen from Fig. 2b, the resistance of hydrothermal $Fe_{1-\delta}Se$ first increases, then decreases with rising temperature, demonstrating a metal to non-metal crossover behavior around $T = 35$ K. Correspondingly, no trace of the resistive kink feature related to the structural/nematic transition is detected. In Fig. 2c, the Hall resistivity $\rho_{xy}(B)$ is proportional to the field at all temperatures, and in Fig. 2d the Hall coefficient $R_H(T)$ displays a non-monotonic variation with temperature. These observations imply the compensated metal character [7-9] down to the lowest measuring temperature. In CVT FeSe, on the other hand, the linear $\rho_{xy}(B)$ is present only at the temperatures above 60 - 75 K, below which significant nonlinearity in $\rho_{xy}(B)$ develops.[7,8] Meanwhile, its positive $R_H(T)$ shows a large drop just below $T_s \sim 90$ K and becomes strongly negative as $T$ is further lowered. All these phenomena are various manifestations of the electron-dominant nematic phase intimately connected with the emergent high-mobility electron band.[7,8] It is also noteworthy that $R_H(T)$ of the present $Fe_{1-\delta}Se$ shows only a small drop at $T \sim 25$ K and its sign keeps positive down to the lowest $T$ (Fig. 2d), indicating a persistent hole-dominant nature. This behavior of $R_H(T)$ for $Fe_{1-\delta}Se$ turns out to be intermediate between those for stoichiometric CVT FeSe [7-9] and hydrothermal $Fe_{1-x}Se$ with a small iron-vacancy $x$,[10] consistent with that $\delta$ is closer to 0.

Therefore, the normal-state electronic properties of our hydrothermal $Fe_{1-\delta}Se$ are distinctly different from those of CVT FeSe. More specifically, no signatures of the structural/nematic transition and the associated high-mobility electron state have been detected. Further, it is interesting that the superconducting $T_c$ (9 K) of the nematic CVT FeSe is lower than that (13.2 K) of the hydrothermal $Fe_{1-\delta}Se$. These results suggest that the highly mobile electron band associated with the nematicity does not favor the superconductivity. Such a $T_c$ suppression in the presence of the nematic order has also been observed among other related systems such as isovalently substituted FeSe systems.[6] Here it is necessary to mention that the higher $T_c$ (~14 K)[20,21] and the electronic nematicity[20] have been reported in the thick films of FeSe, in which additional strain effect from substrate needs to be considered.

Fig. 3a and b present the temperature-dependent resistance $R(T)$ measured under magnetic fields parallel to $c$ axis and $ab$ plane up to 9 T, respectively. Fig. 3c and d show the magnetotransport data $R(B)$ for $B//c$ and $B//ab$, respectively, measured by sweeping the field up to 38.5 T at fixed temperatures. The curves of $R(T)$ in Fig. 3a and b exhibit almost parallel shifts to lower temperatures with increasing magnetic field, *i.e.*, there is no significant field-induced broadening of the resistive transition. This implies a narrow vortex-liquid region in the hydrothermal $Fe_{1-\delta}Se$, as commonly seen in FeSe-11 [15,22-24] and FeAs-122 [25-27] systems.

The data of upper critical field, obtained from the curves of $R(T)$ (Fig. 3a, b) and $R(B)$ (Fig. 3c, d) by the 50 % criteria, are plotted in Fig. 3e. The $H_{c2}(T)$ data display steep slopes near $T_c$, with the values of -7.5 T/K for $B//c$ and around -60 T/K for $B//ab$, and significant downward curvatures at lower temperatures for both $B//c$ and $B//ab$. These indicate the strong orbital-limiting fields and spin paramagnetism[28], respectively. As shown by the solid red/blue curves in Fig. 3e, the data of $H_{c2}(T)$ can be well fitted by the single-band Werthamer-Helfand-Hohenberg (WHH) formula[28] incorporating both the Maki parameter ($\alpha$) and spin-orbit-interaction constant ($\lambda_{so}$). The fittings yield $\alpha = 6$ and $\lambda_{so} = 5$ for $B//c$, and $\alpha = 11.8$ and $\lambda_{so} = 1.5$ for $B//ab$. These parameters reflect the strong spin-paramagnetic effect and spin-orbit interaction in the hydrothermal $Fe_{1-\delta}Se$, similar to isovalently substituted Fe(Se,Te)[22,29,30] and Fe(Te,S)[23] systems. The fitted values of $H_{c2}(0)$ for $B//c$ and $B//ab$ merge with each other at ~42 T, which breaks the BCS weak-coupling Pauli-limit field[31] by a factor of 1.8. By comparison, $H_{c2}(T)$ of CVT FeSe[11,12,14,15] shows a slight downward curvature and breaks the Pauli limit only for $B//ab$, while almost linear in $T$ with a lower $H_{c2}(0)$ (~15 T) below the Pauli limit for $B//c$. The isotropic Pauli-limit-breaking $H_{c2}(0)$ observed in the present $Fe_{1-\delta}Se$ could be explained from the observed steep initial slopes and the fitted parameters, with the significantly larger $\lambda_{so}$ and smaller $\alpha$ for $B//c$. The spin-orbit interaction can reduce the effect of spin paramagnetism thus further enhance $H_{c2}$ [28,32]. We note that similar isotropic $H_{c2}(0)$ has been reported in other iron-based systems having three-dimensional electronic structure.[33,34]

It should also be noted that, although the increased carrier-scattering effect in substituted $FeTe_{0.6}Se_{0.4}$ has been discussed to enhance the initial slope of $H_{c2}$ [29], its value (-13 T/K) of the slope for $B//ab$ is still much smaller than that (around -60 T/K) observed in our $Fe_{1-\delta}Se$. On the other hand, since the non-stoichiometric $\delta$ in $Fe_{1-\delta}Se$ is expected to be as limited as close to CVT FeSe, such a steep initial slope for $B//ab$ of the hydrothermal $Fe_{1-\delta}Se$, which is about one order of magnitude larger than that (-5.6 T/K[12,14,15]) of the CVT FeSe, cannot be accounted for by a very small difference in their iron-vacancy amounts, if any. Combining these and the high sample quality, the unusual large slope of the upper critical field near $T_c$ appears to be an intrinsic property of the present $Fe_{1-\delta}Se$, rather than due to iron-vacancy or disorder scatterings.

To summarize, we have observed unusual electronic and superconducting properties in the hydrothermal $Fe_{1-\delta}Se$. In sharp contrast to the nematic CVT FeSe, our $Fe_{1-\delta}Se$ samples show no signature of the electronic nematicity but an evident metal-nonmetal crossover behavior in the normal state. Meanwhile, a superconducting critical temperature (13.2 K) higher than that (9 K) of nematic FeSe is observed. The present study provides new evidence to prove that the high-mobility electron band associated with the nematicity does not favor the superconductivity. Therefore, broader perspective beyond the electronic nematicity is needed for investigating the origin of the unconventional superconductivity. Furthermore, the upper critical field $H_{c2}(0)$ of the hydrothermal $Fe_{1-\delta}Se$ is found to be isotropic and reach a higher value of ~42 T. This could be understood from the combined effects of the intrinsic strong orbital-limiting field and spin-orbit interaction. The high-quality hydrothermal $Fe_{1-\delta}Se$ can serve as a superior platform for further studies.

This work is supported by the National Key Research and Development Program of China (Grant Nos. 2016YFA0300300 and 2017YFA0303003), the National Natural Science Foundation of China (Grant Nos. 12061131005, 11834016 and 11888101), the Strategic Priority Research Program of Chinese Academy of Sciences (Grant Nos. XDB25000000), and the Strategic Priority Research Program and Key Research Program of Frontier Sciences of the Chinese Academy of Sciences (Grant Nos. QYZDY-SSW-SLH001).

## References

[1] Zhao J, Huang Q, de la Cruz C, Li S L, Lynn J W, Chen Y, Green M A, Chen G F, Li G, Li Z, Luo J L, Wang N L and Dai P C 2008 *Nat. Mater.* **7** 953

[2] Kasahara S, Shi H J, Hashimoto K, Tonegawa S, Mizukami Y, Shibauchi T, Sugimoto K, Fukuda T, Terashima T, Nevidomskyy A H and Matsuda Y 2012 *Nature* **486** 382

[3] Coldea A I and Watson M D 2018 *Annu. Rev. Condens. Matter Phys.* **9** 125

[4] Bohmer A E and Kreisel A 2018 *J. Phys.: Condens. Matter* **30** 023001

[5] Hsu F C, Luo J Y, Yeh K W, Chen T K, Huang T W, Wu P M, Lee Y C, Huang Y L, Chu Y Y, Yan D C and Wu M K 2008 *Proc. Natl. Acad. Sci. USA* **105** 14262

[6] Liu S B, Yuan J, Huh S, Ryu H, Ma M W, Hu W, Li D, Ma S, Ni S L, Shen P P, Jin K, Yu L, Kim C Y, Zhou F, Dong X L and Zhao Z X 2020 *arXiv* 2009.13286

[7] Huynh K K, Tanabe Y, Urata T, Oguro H, Heguri S, Watanabe K and Tanigaki K 2014 *Phys. Rev. B* **90** 144516

[8] Watson M D, Yamashita T, Kasahara S, Knafo W, Nardone M, Beard J, Hardy F, McCollam A, Narayanan A, Blake S F, Wolf T, Haghighirad A A, Meingast C, Schofield A J, Lohneysen H, Matsuda Y, Coldea A I and Shibauchi T 2015 *Phys. Rev. Lett* **115** 027006

[9] Rößler S, Koz C, Jiao L, Rößler U K, Steglich F, Schwarz U and Wirth S 2015 *Phys. Rev. B* **92** 060505(R)

[10] Ni S L, Sun J P, Liu S B, Yuan J, Yu L, Ma M W, Zhang L, Pi L, Zheng P, Shen P P, Li D, Shi D E, Li G B, Sun J L, Zhang G M, Jin K, Cheng J G, Zhou F, Dong X L and Zhao Z X 2019 *arXiv* 1912.12614

[11] Terashima T, Kikugawa N, Kiswandhi A, Choi E S, Brooks J S, Kasahara S, Watashige T, Ikeda H, Shibauchi T, Matsuda Y, Wolf T, Böhmer A E, Hardy F, Meingast C, Löhneysen H v, Suzuki M T, Arita R and Uji S 2014 *Phys. Rev. B* **90** 144517

[12] Audouard A, Duc F, Drigo L, Toulemonde P, Karlsson S, Strobel P and Sulpice A 2015 *Europhys. Lett.* **109** 27003

[13] Farrar L S, Bristow M, Haghighirad A A, McCollam A, Bending S J and Coldea A I 2020 *npj Quantum Mater.* **5** 29

[14] Ok J M, Kwon C I, Kohama Y, You J S, Park S K, Kim J H, Jo Y J, Choi E S, Kindo K, Kang W, Kim K S, Moon E G, Gurevich A and Kim J S 2020 *Phys. Rev. B* **101** 224509

[15] Zhou N, Sun Y, Xi C Y, Wang Z S, Zhang Y F, Xu C Q, Pan Y Q, Feng J J, Meng Y, Yi X L, Pi L, Tamegai T, Xing X Z and Shi Z X 2021 *arXiv* 2102.02353

[16] Ma Y W 2012 *Supercond. Sci. Technol.* **25** 113001

[17] Huang Y L, Feng Z P, Ni S L, Li J, Hu W, Liu S B, Mao Y Y, Zhou H X, Zhou F, Jin K, Wang H B, Yuan J, Dong X L and Zhao Z X 2017 *Chin. Phys. Lett.* **34** 077404

[18] Huang Y L, Feng Z P, Yuan J, Hu W, Li J, Ni S L, Liu S B, Mao Y Y, Zhou H X, Wang H B, Zhou F, Zhang G M, Jin K, Dong X L and Zhao Z X 2017 *arXiv* 1711.02920

[19] Koz C, Schmidt M, Borrmann H, Burkhardt U, Rößler S, Carrillo-Cabrera W, Schnelle W, Schwarz U and Grin Y 2014 *Z. Anorg. Allg. Chem.* **640** 1600

[20] Imai Y, Sawada Y, Nabeshima F, Asami D, Kawai M and Maeda A 2017 *Sci. Rep.* **7** 46653

[21] Feng Z P, Yuan J, Li J, Wu X X, Hu W, Shen B, Qin M Y, Zhao L, Zhu B Y, Stanev V, Liu M, Zhang G M, Yang H X, Li J Q, Dong X L, Zhou F, Zhou X J, Kusmartsev F V, Hu J P, Takeuchi I, Zhao Z X and Jin K 2018 *arXiv* 1807.01273

[22] Lei H C, Hu R W, Choi E S, Warren J B and Petrovic C 2010 *Phys. Rev. B* **81** 094518

[23] Lei H C, Hu R W, Choi E S, Warren J B and Petrovic C 2010 *Physical Review B* **81** 184522

[24] Yuan D N, Huang Y L, Ni S L, Zhou H, Mao Y Y, Hu W, Yuan J, Jin K, Zhang G M, Dong X L and Zhou F 2016 *Chin. Phys. B* **25** 077404(R)

[25] Wang Z S, Luo H Q, Ren C and Wen H H 2008 *Phys. Rev. B* **78** 140501(R)

[26] Bukowski Z, Weyeneth S, Puzniak R, Moll P, Katrych S, Zhigadlo N D, Karpinski J, Keller H and Batlogg B 2009 *Phys. Rev. B* **79** 104521

[27] Wang Z S, Xie T, Kampert E, Förster T, Lu X Y, Zhang R, Gong D L, Li S L, Herrmannsdörfer T, Wosnitza J and Luo H Q 2015 *Phys. Rev. B* **92** 174509

[28] Werthamer N R, Helfand E and Hohenberg P C 1966 *Phys. Rev.* **147** 295

[29] Khim S, Kim J W, Choi E S, Bang Y, Nohara M, Takagi H and Kim K H 2010 *Phys. Rev. B* **81** 184511

[30] Tarantini C, Gurevich A, Jaroszynski J, Balakirev F, Bellingeri E, Pallecchi I, Ferdeghini C, Shen B, Wen H H and Larbalestier D C 2011 *Phys. Rev. B* **84** 184522

[31] Clogston A M 1962 *Phys. Rev. Lett.* **9** 266

[32] Klemm R A, Luther A and Beasley M R 1975 *Phys. Rev. B* **12** 877

[33] Yuan H Q, Singleton J, Balakirev F F, Baily S A, Chen G F, Luo J L and Wang N L 2009 *Nature* **457** 565

[34] Fang M H, Yang J H, Balakirev F F, Kohama Y, Singleton J, Qian B, Mao Z Q, Wang H D and Yuan H Q 2010 *Phys. Rev. B* **81** 020509(R)

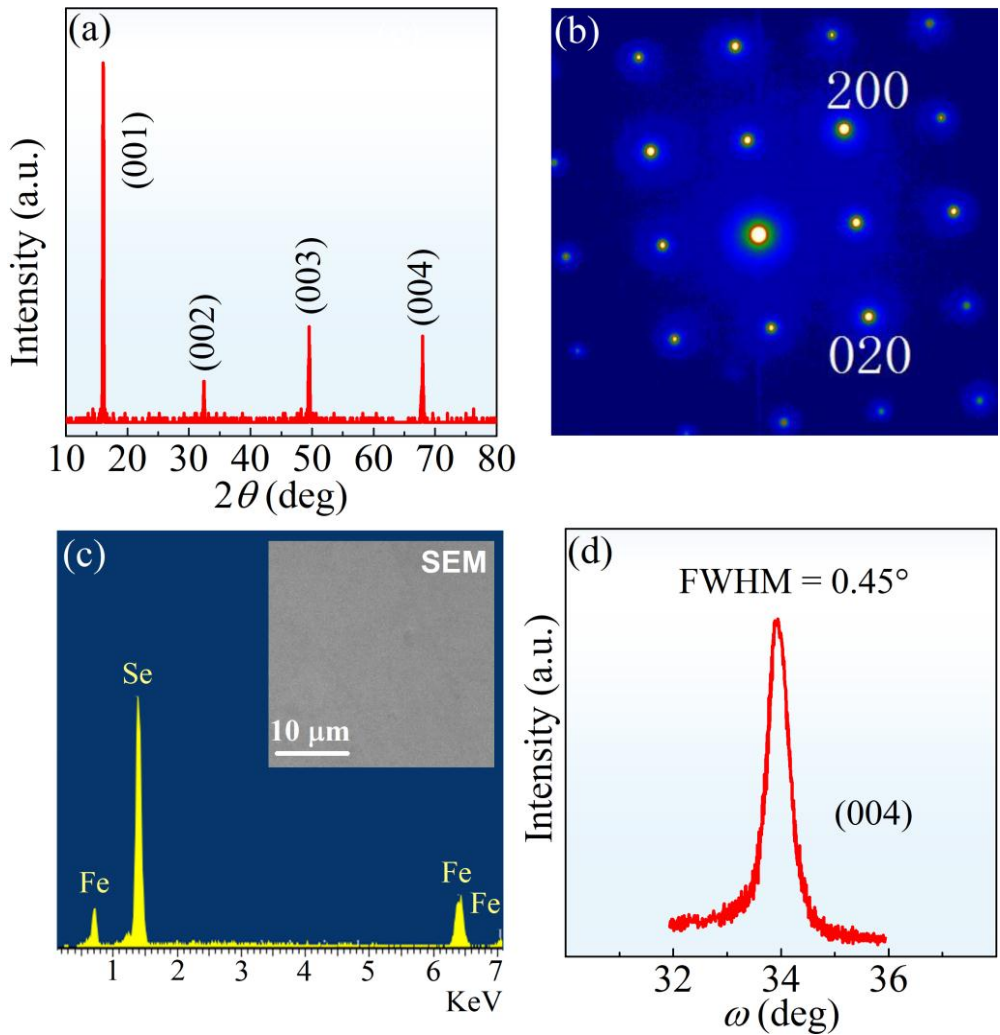

Fig. 1. (a) The XRD pattern of the hydrothermal flake-form $Fe_{1-\delta}Se$ shows a single preferred (00l) orientation. (b) The SAED pattern along [00l] zone axis. Both the XRD and SAED patterns are consistent with the crystallographic symmetry of *P*4/*nmm*. (c) No impurities, but only the Fe and Se components, are identified by EDX spectroscopy. Inset: A smooth surface morphology of the sample is revealed by SEM image. (d) The double-crystal XRD rocking curve for (004) reflection gives an FWHM = 0.45°, demonstrating a high sample quality.

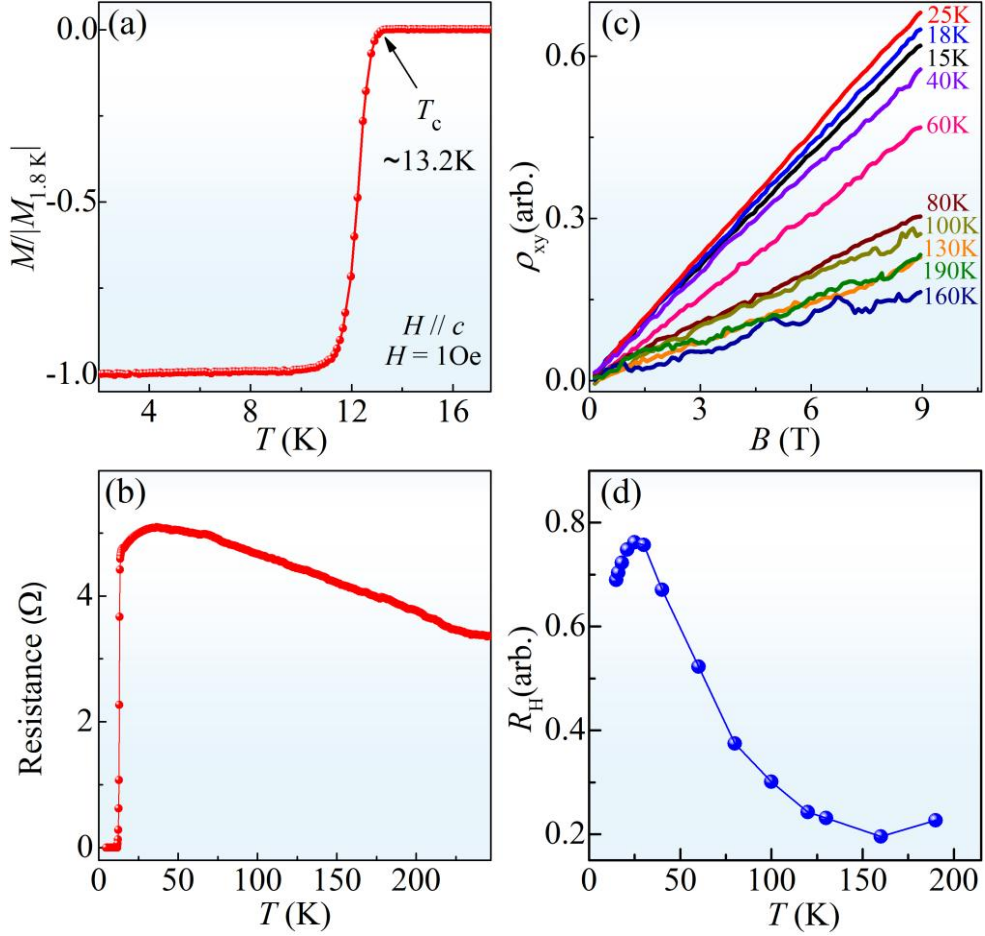

Fig. 2. (a) The normalized diamagnetic curve of the hydrothermal $Fe_{1-\delta}Se$. (b) The temperature-dependent resistance is maximized at $T \sim 35$ K in the normal state. (c) The Hall resistivity $\rho_{xy}(B)$ is proportional to field $B$ at all the temperatures. (d) The non-monotonic temperature dependence of Hall coefficient $R_H$, obtained from the slope of $\rho_{xy}(B)$, shows a small drop below $T \sim 25$ K but no sign change.

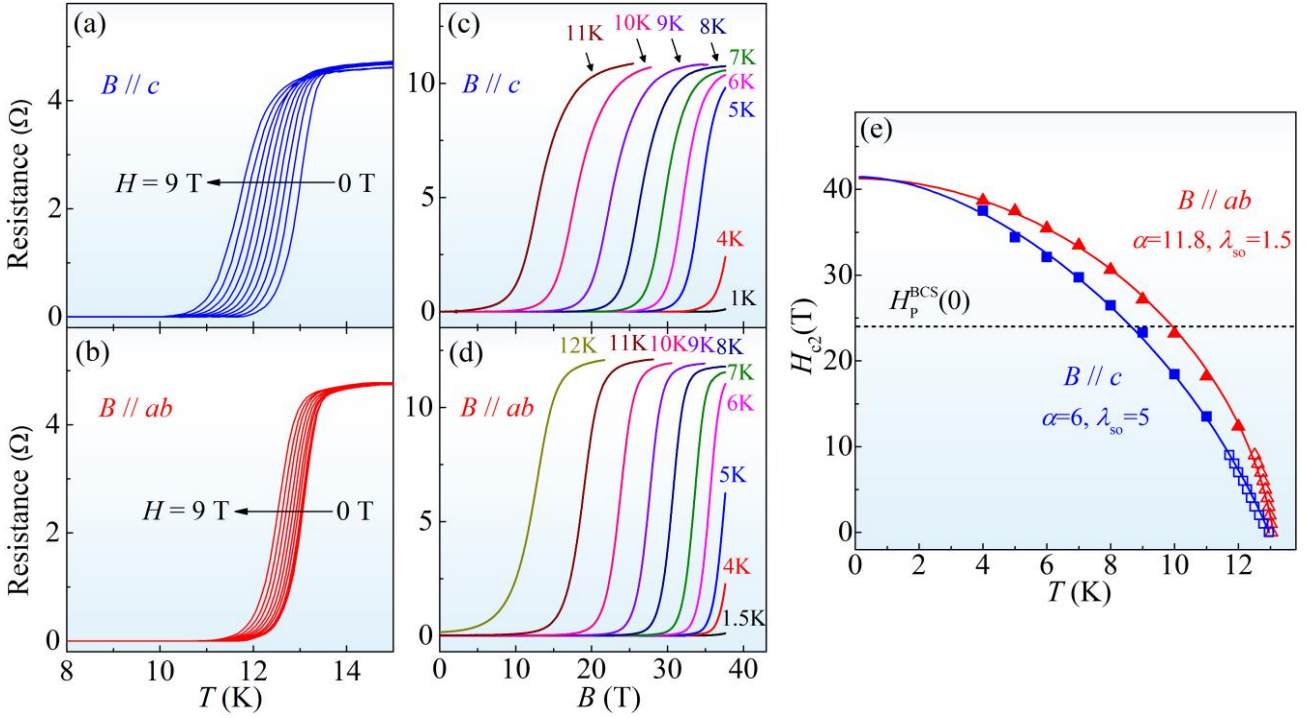

Fig. 3. (a) and (b) are the $R(T)$ curves of the hydrothermal $Fe_{1-\delta}Se$ for $B//c$ and $B//ab$ up to 9 T, respectively. (c) and (d) show the $R(B)$ curves for $B//c$ and $B//ab$, respectively, measured by sweeping the field up to 38.5 T at fixed temperatures. (e) The $H_{c2}(T)$ data for $B//c$ (closed or open squares) and $B//ab$ (closed or open triangles), obtained from the $R(T)$ (the open symbols) and $R(B)$ (the closed symbols) data at 50% of the normal-state resistance. The high-field data below 6 K are obtained by linear extrapolation. The slopes of $H_{c2}(T)$ near $T_c$ are around -60 T/K for $B//ab$ and -7.5 T/K for $B//c$. The solid red/blue curves are the results of single-band WHH fitting with both the spin-paramagnetic effect ($\alpha$) and spin-orbit interaction ($\lambda_{so}$) introduced. The BCS Pauli-limit field ($H_P^{BCS}(0)=1.84T_c \sim 24$ T) is marked by the dotted horizontal line.